# Effects of carbon pricing and other climate policies on $CO_2$ emissions

Emanuel Kohlscheen [a], Richhild Moessner [a, b] and Előd Takáts [a]

**Abstract**

We provide ex-post empirical analysis of the effects of climate policies on carbon dioxide emissions at the aggregate national level. Our results are based on a comprehensive database of 121 countries. As climate policies we examine carbon taxes and emissions trading systems (ETS), as well as the overall stringency of climate policies. We use dynamic panel regressions, controlling for macroeconomic factors such as economic development, GDP growth, urbanisation, as well as the energy mix. We find that higher carbon taxes and prices of permits in ETS reduce carbon emissions. An increase in carbon taxes by $10 per ton of $CO_2$ reduces $CO_2$ emissions per capita by 1.3% in the short run and by 4.6% in the long run.

[a] Bank for International Settlements, Centralbahnplatz 2, 4002 Basel, Switzerland. *E-mail addresses*: emanuel.kohlscheen@bis.org; richhild.moessner@bis.org; elod.takats@bis.org. The views expressed in this paper are those of the authors and do not necessarily reflect those of the Bank for International Settlements.

[b] CESifo, Munich, Germany and National Institute of Economic and Social Research, London, United Kingdom.

# 1. Introduction

Emissions of carbon dioxide ($CO_2$), a key driver of global warming and climate change, have continued to increase globally in recent years. If current climate policies are unchanged, standard climate change scenarios predict an increase of 3 degrees Celsius (°C) in global temperatures compared with pre-industrial levels by the end of the century (10, 13). This could have catastrophic consequences.[1] To avoid such a scenario, climate policies need to be expanded in order to reduce the emission of carbon dioxide and other greenhouse gases. Such policies include carbon taxes and emissions trading systems (ETS), but also broader changes in regulation (30, 31).

To be consistent with global emissions that limit an overshoot of the goal from the Paris agreement of 1.5°C global warming above pre-industrial levels, global net anthropogenic $CO_2$ emissions need to decline by about 45% from their 2010 level by 2030, reaching net zero around 2050. To cap global warming at 2°C, $CO_2$ emissions need to decrease by about 25% from the 2010 level by 2030 and reach net zero around 2070. Yet, according to latest estimates, the total global greenhouse gas emission level in 2030 is expected to be 16% above the 2010 level (33).

Carbon pricing can be an effective policy to reduce emissions. Higher carbon prices make low-carbon energy more competitive, provide incentives to reduce emissions, and reduce demand for carbon-intensive fuels (1, 18, 23). Moreover, a strong commitment to higher carbon prices by governments provides incentives for investors to invest in the expansion and development of low-carbon technologies (23). Yet, the aggregate impact of these regulations remains unclear.

This paper analyses the effects of climate policies on $CO_2$ emissions. We provide ex-post empirical analysis of the effects of carbon pricing on carbon emissions at the aggregate national level, based on a comprehensive database covering 121 countries. We rely on carbon emissions data and macroeconomic variables over the 1971-2016 period, as well as data on climate policies. As climate policies we consider national and supranational carbon taxes and emissions trading systems (ETS). We also consider a broad index that measures the overall stringency of climate policies for OECD and major emerging economies. This index captures regulatory responses that go beyond carbon pricing. We use dynamic panel regressions, to account for the large degree of persistence in emissions, and control for macroeconomic factors such as economic development, GDP growth, urbanisation, as well as the composition of the electricity mix based on (15).

Overall, we find that higher carbon taxes and prices of permits in ETS have significantly reduced carbon emissions. An increase in carbon taxes by $10 per ton of $CO_2$ equivalents ($tCO_2$) reduces $CO_2$ emissions per capita by 1.3% in the short run and by 4.6% in the long run. This effect is statistically significant in all econometric specifications, with *p-values* that are always below 0.01. The same increase in the prices of ETS permits also reduces $CO_2$ emissions per capita by 1.4% in the short run and 5.0% in the long run. The magnitude of this effect however varies more across specifications and is statistically significant in five out of nine specifications (with *p-values* below 0.05). Further, more stringent climate policies as measured by a broad index for OECD and major emerging economies also significantly reduces carbon emissions, with a standard deviation increase in the index reducing $CO_2$ emissions per capita by around 1.5% in the short run and 6% in the long run (*p-values* below 0.05 in five out of six specifications). The estimates are robust to inclusion of controls for the overall quality of governance in the respective countries, as proxied by an index for control of corruption.

---

[1] Climate change due to increases in carbon dioxide concentrations could be irreversible for 1,000 years after emissions stop (32).

The paper adds to the growing literature that examines the impact of carbon taxes and ETS on $CO_2$ emissions - reviewed recently in (9). As she notes, relatively few papers analyse the *ex-post* effects of carbon pricing on carbon emissions at the aggregate national level. And those papers that do, focus on a few specific countries and then extrapolate the results to a broader set of countries. We contribute to this literature by analysing a wide range of countries. This allows us to obtain more precise estimates of the impact of policies.

Our findings are relevant for future climate policies. More specifically, the findings that higher carbon taxes and prices of permits in ETS have reduced carbon emissions provide evidence that such tools could speed up the necessary transition to a world with much lower emissions. Additionally, our findings suggest that policymakers can rely on a wider range of climate policies to speed up the transition to lower carbon emissions, ideally to net zero emissions.

The remainder of the paper proceeds as follows. Section 2 discusses the related literature. Section 3 summarises the data. Section 4 presents the methodology. Section 5 presents our empirical estimates. Section 6 concludes.

## 2. Literature

The paper mainly builds on two strands of the literature. First, it builds on the literature examining the effects of climate policies, such as carbon taxes and ETS, on carbon emissions. Second, it builds on the literature studying macroeconomic determinants of carbon emissions.

The effects of climate policies, such as carbon taxes and ETS, on carbon emissions have been studied in several papers. A comprehensive recent review of the literature on the effects of carbon pricing on carbon emissions is (9). She highlights that surprisingly few papers have conducted an *ex-post* empirical analysis of how carbon pricing has actually affected $CO_2$ emissions,[2] and that the vast majority of these papers are focused on Europe. She notes that in most cases, studies have estimated emissions reductions in the sectors covered by the carbon pricing policy, although some extrapolate to broader jurisdictional effects (3, 22, 26).[3] The author concludes that the majority of the studies suggest that the aggregate reductions from carbon pricing on emissions are limited, generally between 0% and 2% per year, with considerable variation across sectors. She also concludes that in general carbon taxes perform better than ETS; and that studies of the European Union's ETS indicate limited average annual reductions in carbon emissions of 0% to 1.5%.

Our paper contributes to filling this gap in the literature by providing *ex-post* empirical analysis of the effects of carbon pricing on carbon emissions at the aggregate national level, for a very broad sample of countries. There is also little empirical literature on the effects of broader climate policies, and we contribute to fill this gap by studying the effects of an index of broad climate policies on $CO_2$ emissions.

---

[2] The author notes that this is despite a large theoretical literature on carbon pricing, often using model simulations, predictive models or theoretical assessments of reductions, with these prospective analyses constituting the vast majority of the quantitative literature on carbon pricing (9).
[3] The effects of carbon prices in five sectors for a large number of countries since the 1990s are studied in (26). They find that the introduction of carbon pricing reduced growth in total $CO_2$ emissions by 1%-2.5% on average relative to imputed counterfactuals, with the greatest reduction in the electricity and heat sector.

There is limited empirical evidence that higher carbon prices reduce carbon emissions (1, 18, 23). For OECD economies an increase in a broad-based tax on energy consumption of €10/tCO$_2$ is expected to lead to a 7.3% reduction in carbon emissions from fossil fuel consumption in the long run (29). Europe's carbon taxes led to a cumulative reduction of around 4% to 6% for a \$40/tCO$_2$ tax covering 30% of emissions (21). The authors argue that emissions reductions would likely be larger for a broad-based US carbon tax, since European carbon taxes do not include sectors with the lowest marginal costs of carbon pollution abatement in the tax base. Based on sectoral analysis, an imprecisely estimated semi-elasticity of a 0.5% reduction in carbon emissions growth per average \$10/tCO$_2$ carbon price is found in (26).[4]

The effects of macroeconomic variables on carbon emissions have been studied in (5, 7, 8, 12, 14, 19, 24, 27, 28, 34). Our approach for controlling for macroeconomic determinants of carbon emissions follows (15). For a review of the literature studying the macroeconomic determinants of carbon emissions readers are referred to that study.

A review of ex-post empirical assessments on the impact of carbon pricing on competitiveness in OECD and G20 countries in the electricity and industrial sectors concludes that there are no significant effects on competitiveness (6). In turn, (16) review the empirical evidence available in academic ex-post analyses of the effectiveness of carbon pricing schemes in promoting technological change necessary for full decarbonisation in the cases of the European Union, New Zealand and Scandinavia. They find that there is little empirical evidence so far, with most of the papers on the topic being theoretical. Recent papers studying the macroeconomic effects of carbon taxes include (20, 21).

## 3. Data

In order to assess the empirical drivers of carbon dioxide emissions, we collect data from several sources. Data on carbon dioxide (CO$_2$) emission per capita, measured in metric tons, are from the Carbon Dioxide Information Analysis Center (CDIAC) at the U.S. Department of Energy's Oak Ridge National Laboratory. Demographic, economic and energy use data are from the World Bank (GDP per capita, GDP growth, the urbanisation rate, the share of manufacturing in GDP and the share of coal, oil and renewables in electricity generation).[5] Our database spans the 1971-2016 period and 121 countries. The country selection is based solely on data availability.

Our main variables of interest are CO$_2$ emissions and climate policies.

Carbon emissions emanate mostly from the burning of fossil fuels (to heat, transport goods and people or generate electricity) and the manufacture of steel and cement. The level of carbon emissions, following that of economic development, is highly uneven across countries. Highly developed advanced economies tend to emit large quantities of carbon per capita, while less developed economies, particularly in Africa, tend to emit less (Figure 1). However, there is no definite advanced-emerging economy divide: due to fast economic development many emerging market economies (including energy producers and fast developing East Asian economies) already have high carbon emission levels per capita. Besides, given their larger populations, their contribution to global CO$_2$ emissions has grown very rapidly.

---

[4] See (9) for further references.
[5] Note that we include only the respective shares of the energy mix, and not the intensity of use.

**Carbon emissions per capita in 2016**

Figure 1

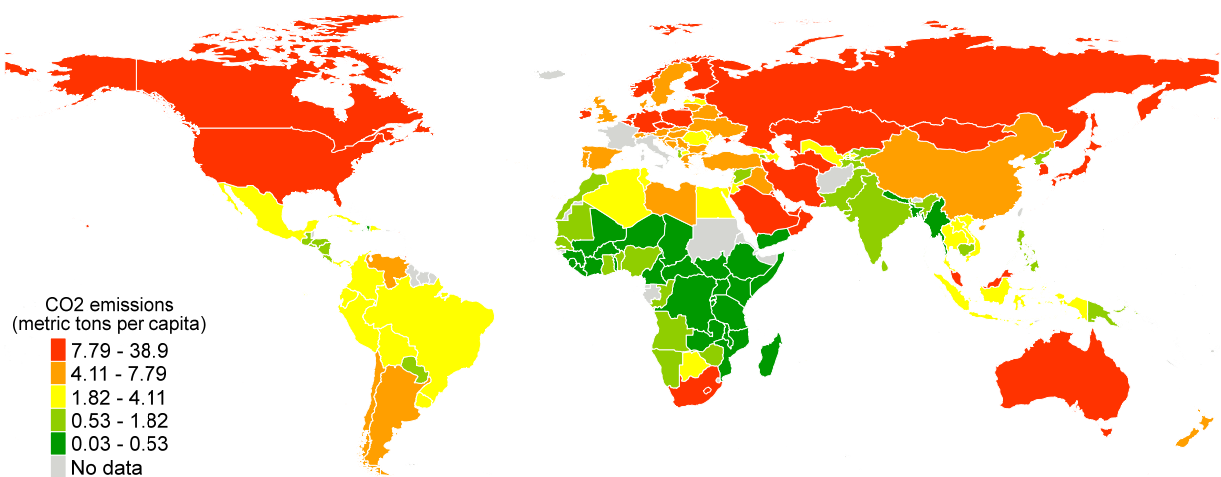

Primary climate policies are carbon taxes and ETS. The first carbon tax was introduced in Finland in 1990. In turn, emissions trading has been considered a possible tool for mitigating greenhouse gas emissions since the early 1990s and formed a key part of the Kyoto Protocol agreement (25). Our database contains information on carbon taxes and ETS implemented at the national and supranational levels since 1990 from the World Bank (35). Country coverage and prices of these policies are shown in Figures 2 and 3 separately for carbon taxes and ETS, for the end of our sample period in 2016. The share of global greenhouse gas emissions covered by the carbon taxes and ETS at the national level is shown in Figure 4. As the Figure shows, there has been a large increase in this proportion.[6]

In the case of carbon taxes, governments set the price of carbon emissions, and let private agents determine emissions reductions. ETS have two main forms, cap-and-trade and baseline-and-credit ETS. For cap-and-trade ETS, governments set a limit on emissions, and allowances up to this limit are auctioned or allocated according to certain criteria. These permits are then traded and carbon prices are determined by supply and demand in the market. For baseline-and-credit ETS, baselines for emissions are set for regulated emitters. Emitters with emissions above their designated baseline need to give up credits to make up for these emissions, while those with emissions below their baseline receive credits for these reductions, which they can sell to other emitters (35).

---

[6] Coverage and prices for the latest available data in 2020 are shown in Appendix Figures A1 and A2, from which we can see that the coverage of these policies across countries has increased further up to 2020.

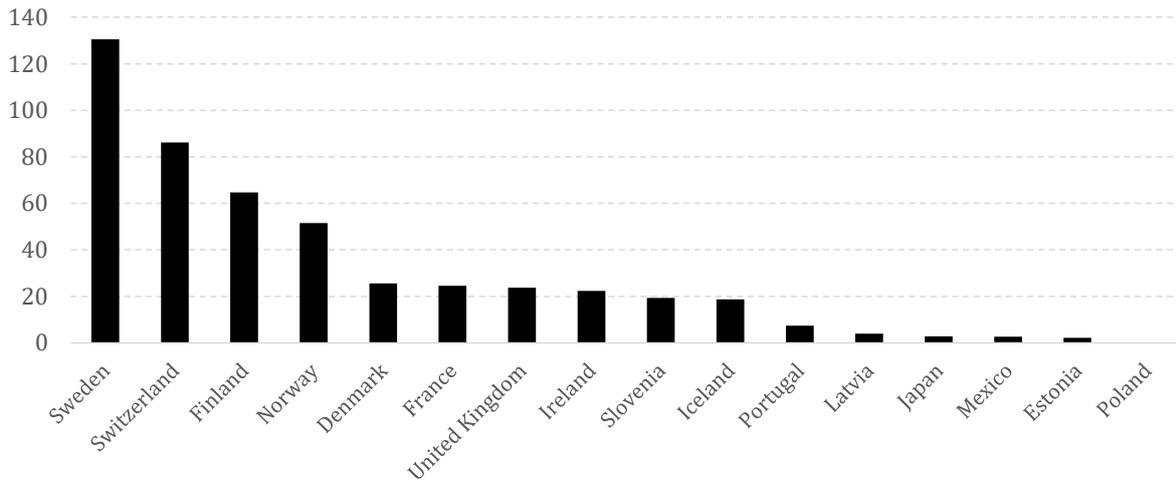

Figure 2: Prices of cabon taxes as of 2016

Note: Nominal prices as of 1 April 2016 in $/tCO2 equivalents. In the case of the United Kingdom, the number refers to the carbon price floor.

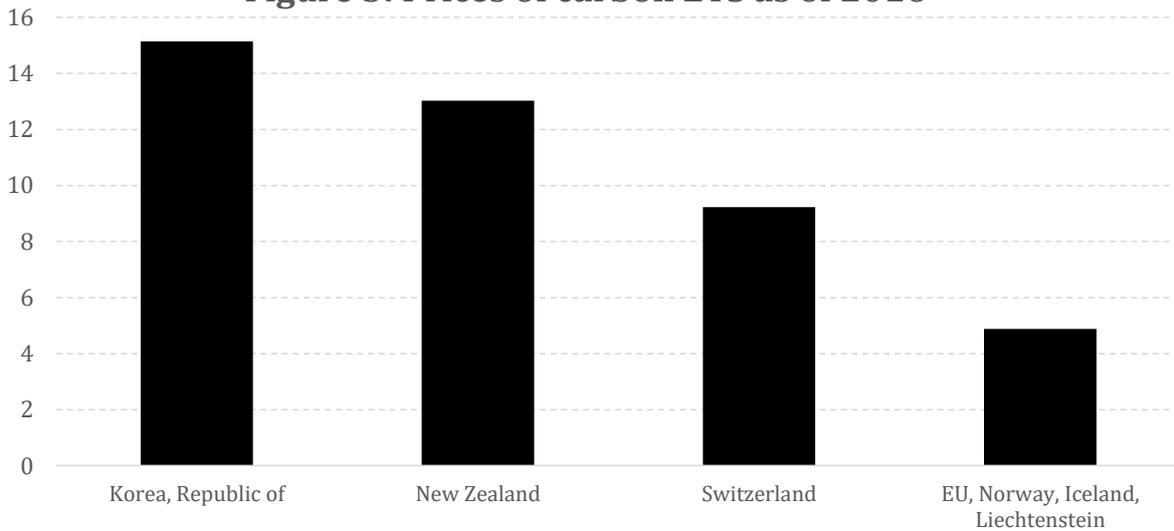

Figure 3: Prices of carbon ETS as of 2016

Note: Nominal prices as of 1 April 2016 in $/tCO2 equivalents. EU: European Union.

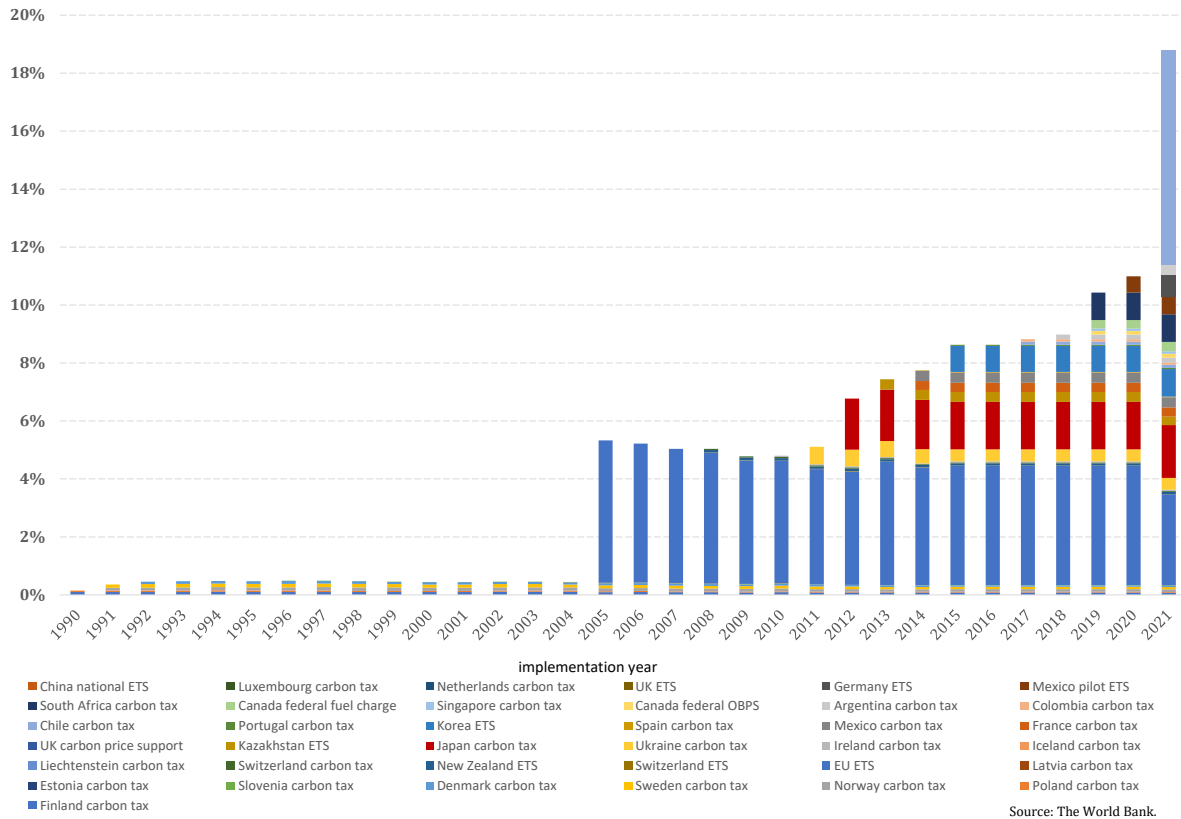

Figure 4: Global greenhouse gas emissions covered by carbon taxes and ETS at the national and supranational levels

Source: The World Bank.

Besides carbon taxes and ETS, we also evaluate the effectiveness of climate policies by using a broader index for the overall stringency of climate policies for OECD member countries and major emerging economies. This is based on the Environmental Policy Stringency Index (EPS), which is compiled by the OECD and available from 1990 to 2015. The index covers a much wider range of climate policies, also considering other regulatory policies (4). Generally, stringency is defined as the degree to which environmental policies put an explicit or implicit price on polluting or, more generally, on environmentally harmful behaviour.[7]

We conduct several checks of the robustness of our baseline results. In particular, to control for the quality of governance in a country, we use a measure of the control of corruption from the World Bank. It reflects perceptions of the extent to which public power is exercised for private gain, including both petty and grand forms of corruption, as well as "capture" of the state by elites and private interests.[8]

---

[7] The index ranges from 0 (not stringent) to 6 (highest degree of stringency). It covers 28 OECD and 6 BRIICS countries (Brazil, China, India, Indonesia, Russia, South Africa) for the period 1990-2012, and has been extended for a subset of these countries until 2015 (France, Germany, Italy, Japan, Korea, United Kingdom, United States, Brazil, China, India, Indonesia, Russia, South Africa). The EPS index is based on the degree of stringency of 14 environmental policy instruments, primarily related to climate and air pollution.

[8] This indicator ranges from approximately -2.5 (weak) to 2.5 (strong) governance performance.

## 4. Methodology

We quantify the effects of climate policies on carbon emissions through a dynamic panel model. The dynamic specification accounts for the high degree of persistence in $CO_2$ emissions. Throughout, we control for macroeconomic factors, such as economic development (GDP per capita, GDP growth, urbanisation, the share of manufacturing in total output, as well as the energy mix used in electricity production.[9] Formally, we explore the effects of climate policies (denoted by CP) on the logarithm of carbon-dioxide emissions per capita (denoted by lnCO2) according to:

$$\ln CO2_{i,t} = \alpha_i + \beta_t + \rho \ln CO2_{i,t-1} + \lambda\, CP_{i,t-1} + \gamma\, GDPpc_{i,t} + \theta\, growth_{i,t} \\ + \omega\, urbanisation_{i,t} + \vartheta\, manufacturing_{i,t} + \mu_1 share_{oil,i,t} \\ + \mu_2\, share_{coal,i,t} + \mu_3\, share_{renewables,i,t} + \varepsilon_{i,t} \tag{1}$$

Besides the lagged dependent variable, we include climate policies as key explanatory variables of interest. To address potential endogeneity concerns, we use lagged climate policy variables, which minimise the risk of reverse causality. As climate policies we consider carbon taxes and prices of permits in ETS implemented at the national level, as well as the broader index for the stringency of climate policies described in Section 2. We also include a number of macroeconomic variables as controls.

Throughout, our models include country fixed effects to capture unobserved heterogeneities across countries that might affect the rate of carbon dioxide emissions. These include fixed institutional factors like enforcement of environmental laws as well as natural factors such as average median temperatures, which tend to correlate with heating or cooling needs. We also include the full set of yearly time dummies to control for the effects of global factors. These subsume for instance technological advances that may reduce environmental effects, as well as other global trends or global shocks.

Throughout we use fixed effect panel estimations and base inference on cluster robust standard errors. As a robustness check, we control additionally for the quality of governance in a country, using the measure of the control of corruption as described in Section 3.

---

[9] Note that we include only the respective shares of the energy mix, and not the intensity of use.

## 5. Estimation results

### 5.1 Effects of climate policies

We first present baseline estimates of our dynamic panel equation (1). The estimations for the effects of carbon taxes, prices of permits in the ETS and the EPS policy index separately are shown in columns I, II and IV of Table 1, respectively. The results when including carbon taxes and prices of permits of ETS together are shown in column III, and when including all three climate policies together in column V.

Overall, the specifications describe the country-specific evolution of carbon emissions quite well. Our models with climate policies are able to explain 97–98% of the variation in per capita $CO_2$ emissions across countries and across time (Table 1). Importantly, this is driven by variation both within and between countries, as the $R^2$ within ranges from 0.82 to 0.88, and the $R^2$ between ranges from 0.97 to 0.99. The lagged dependent variable has a coefficient of just above 0.7 and is clearly statistically significant - which confirms that a dynamic panel specification is indeed appropriate. Put differently, more that 70% of $CO_2$ emissions can be explained by previous year emissions alone.[10] Most of the macroeconomic control variables are statistically significant with the expected signs.

Consistently we find that higher carbon taxes significantly reduce carbon emissions (*p-values* always below 0.01). On average, an increase in carbon taxes by $10 per ton of $CO_2$ ($tCO_2$) reduces $CO_2$ emissions per capita by 1.3% in the short run and by 4.6% in the long run (baseline model in column III of Table 1).[11] Importantly, this result is robust to controlling for ETS prices (see column III).

We also find that higher prices of ETS permits reduce carbon emissions (*p-values* below 0.01 in columns I and III of Table 1). An increase in prices of ETS permits by $10/tCO_2$ reduces $CO_2$ emissions per capita by 1.4% in the short run and by 5.0% in the long run, when carbon taxes are controlled for (column III).

Moreover, we find that the broad EPS index of climate policy stringency has a negative effect on carbon emissions at the 10% significance level (column IV of Table 1). A one standard deviation increase in the EPS index reduces $CO_2$ emissions per capita by 1.6% in the short run and 6.2% in the long run. It is striking that these results are broadly similar to the results we obtain for carbon taxes, for which a standard deviation increase leads to a reduction in carbon emissions of 1.1% in the short run and 3.9% in the long run.

---

[10] Of course, the high fraction of variation that is explained by the model is partly related to this high degree of persistence in $CO_2$ emissions. This could be a result of strong habit persistence of consumers, or the persistent nature of many industrial production processes.

[11] The long-run effect is $\lambda/(1-\rho)$.

**Table 1**
**Effects of climate policies on CO2 emissions**

Dependent variable: lnCO2 emissions per capita (in metric tons, log)

| | model | | | | |
|---|---|---|---|---|---|
| | (I) | (II) | (III) | (IV) | (V) |
| | model with carbon tax | model with ETS | model with carbon tax and ETS | model with EPS | model with carbon tax, ETS and EPS |
| previous year lnCO2 per capita | 0.7206*** | 0.7231*** | 0.7192*** | 0.7471*** | 0.7334*** |
| | 0.0541 | 0.0538 | 0.0542 | 0.0598 | 0.0627 |
| carbon tax (lagged) | -0.0014*** | | -0.0013*** | | -0.0006*** |
| | 0.0003 | | 0.0003 | | 0.0002 |
| ETS (lagged) | | -0.0019*** | -0.0014*** | | -0.0003 |
| | | 0.0005 | 0.0005 | | 0.0004 |
| EPS (lagged) | | | | -0.0164** | -0.0135** |
| | | | | 0.0064 | 0.0066 |
| GDP per capita (log) | 0.1877*** | 0.1854*** | 0.1893*** | 0.1712*** | 0.1801*** |
| | 0.0436 | 0.0432 | 0.04 | 0.0575 | 0.0616 |
| GDP growth | 0.4050*** | 0.3982*** | 0.3946*** | 0.5099*** | 0.5012*** |
| | 0.0982 | 0.0979 | 0.0977 | 0.1334 | 0.1334 |
| urbanisation rate | 0.3615** | 0.3643** | 0.3519** | 0.2500* | 0.2086 |
| | 0.1632 | 0.1622 | 0.1623 | 0.1396 | 0.1352 |
| manufacturing / GDP | 0.5435*** | 0.5450*** | 0.5445*** | 0.1661 | 0.1036 |
| | 0.1401 | 0.1404 | 0.14 | 0.1875 | 0.2038 |
| share of electricity from oil | 0.1571*** | 0.1530*** | 0.1565*** | 0.1387** | 0.1614** |
| | 0.0396 | 0.0392 | 0.0396 | 0.0564 | 0.0592 |
| share of electricity from coal | 0.2415*** | 0.2348*** | 0.2334*** | 0.2331*** | 0.2640*** |
| | 0.0734 | 0.0720 | 0.0719 | 0.0642 | 0.0747 |
| share of electricity from renewables | -0.1836** | -0.1831** | -0.1589** | -0.2361** | -0.1975** |
| | 0.0782 | 0.0792 | 0.075 | 0.0858 | 0.0852 |
| observations | 3881 | 3881 | 3881 | 653 | 653 |
| number of countries | 121 | 121 | 121 | 30 | 30 |
| R2 within | 0.822 | 0.822 | 0.822 | 0.879 | 0.881 |
| R2 between | 0.989 | 0.989 | 0.989 | 0.971 | 0.971 |

Note: Sample period: from 1971 to 2016, annual data. Cluster-robust standard errors reported in brackets are clustered at the country level. \*\*\*/\*\*/\* denote statistical significance at the 1%/5%/10% level. (I): nominal price of first carbon tax in USD/tCO2 equivalents; (II): nominal price of ETS in USD/tCO2 equivalents; (IV): EPS index.

When we include all three measures together, the coefficients on carbon taxes and the EPS index remain significantly negative and have a somewhat smaller magnitude (column V). While the coefficient on the prices of ETS permits remains negative, it loses significance. This is most likely due to the much smaller sample size, as we lose more than 80% of the sample when including the EPS index. Further, the EPS index also reflects carbon prices, which introduces the problem of multicollinearity of the variables – which is particularly acute for small sample sizes.

## 5.2 Robustness

We perform several robustness checks. First, we also control for the quality of governance in a country, proxied by a measure of control of corruption described in Section 3 (Table 2).[12] We find that higher carbon taxes and more stringent climate policy based on the broader EPS policy index significantly reduce carbon emissions when controlling for governance in all specifications (Table 2). By contrast, the coefficient on the prices of ETS permits remains significant only in one specification (column II). These results imply that the effects of carbon taxes and the broad climate policy index on carbon emissions are generally more robust than those for the effects of ETS permit prices.

---
[12] Note that corruption could for instance undermine the effectiveness of environmental regulations.

**Table 2**
**Effects of climate policies on CO2 emissions: with control of corruption**
Dependent variable: lnCO2 emissions per capita (in metric tons, log)

|  | model | | | | |
|---|---|---|---|---|---|
|  | (I) | (II) | (III) | (IV) | (V) |
|  | model with carbon tax | model with ETS | model with carbon tax and ETS | model with EPS | model with carbon tax, ETS and EPS |
| previous year lnCO2 per capita | 0.7300*** | 0.7328*** | 0.7297*** | 0.6265*** | 0.6110*** |
|  | 0.0257 | 0.0255 | 0.0257 | 0.0627 | 0.0643 |
| carbon tax (lagged) | -0.0011*** |  | -0.0011*** |  | -0.0007*** |
|  | 0.0003 |  | 0.0003 |  | 0.0003 |
| ETS (lagged) |  | -0.0006** | -0.0004 |  | -0.0002 |
|  |  | 0.0003 | 0.0003 |  | 0.0005 |
| EPS (lagged) |  |  |  | -0.0168** | -0.0154* |
|  |  |  |  | 0.0072 | 0.0077 |
| GDP per capita (log) | 0.2006*** | 0.2030*** | 0.2007*** | 0.2954*** | 0.3060*** |
|  | 0.0357 | 0.0360 | 0.0358 | 0.0568 | 0.0590 |
| GDP growth | 0.4738*** | 0.4646*** | 0.4677*** | 0.3402*** | 0.3258*** |
|  | 0.1068 | 0.1061 | 0.1064 | 0.1015 | 0.1013 |
| urbanisation rate | 0.2689 | 0.2754 | 0.2595 | 0.2527 | 0.1735 |
|  | 0.1778 | 0.1767 | 0.1778 | 0.2147 | 0.2092 |
| manufacturing / GDP | 0.2968* | 0.3068* | 0.2927* | 0.3496 | 0.1882 |
|  | 0.1582 | 0.1595 | 0.1591 | 0.2908 | 0.2917 |
| share of electricity from oil | 0.1299*** | 0.1272*** | 0.1297*** | 0.2410*** | 0.2711*** |
|  | 0.0430 | 0.0428 | 0.0431 | 0.0812 | 0.0862 |
| share of electricity from coal | 0.2848*** | 0.2779*** | 0.2816*** | 0.3427*** | 0.3767*** |
|  | 0.0637 | 0.0642 | 0.0643 | 0.0949 | 0.1047 |
| share of electricity from renewables | -0.2730*** | -0.2833*** | -0.2669*** | -0.2210** | -0.1984* |
|  | 0.0801 | 0.0830 | 0.0797 | 0.1018 | 0.0980 |
| control of corruption (lagged) | -0.0032 | -0.0035 | -0.0032 | 0.0149 | 0.0179 |
|  | 0.0185 | 0.0184 | 0.0185 | 0.0173 | 0.0170 |
| observations | 1845 | 1845 | 1845 | 441 | 441 |
| number of countries | 119 | 119 | 119 | 30 | 30 |
| R2 within | 0.778 | 0.778 | 0.778 | 0.855 | 0.858 |
| R2 between | 0.988 | 0.987 | 0.988 | 0.902 | 0.901 |

Note: Sample period: from 1971 to 2016, annual data. Cluster-robust standard errors reported in brackets are clustered at the country level. ***/**/* denote statistical significance at the 1%/5%/10% level. (I): nominal price of first carbon tax in USD/tCO2 equivalents; (II): nominal price of ETS in USD/tCO2 equivalents; (IV): EPS index.

Furthermore, we also estimate the effects of carbon policies when using real carbon prices instead of nominal ones. That is, we consider prices obtained by deflating the nominal USD carbon price by the US consumer price index (Table 3). The results for the significantly negative effects of higher carbon taxes, of higher prices of ETS permits and of more stringent climate policy based on the EPS index on $CO_2$ emissions are robust to using these alternative measures.

**Table 3**
**Effects of climate policies on CO2 emissions: using real carbon prices**

Dependent variable: lnCO2 emissions per capita (in metric tons, log)

| | model | | | | |
|---|---|---|---|---|---|
| | (I) | (II) | (III) | (IV) | (V) |
| | model with carbon tax | model with ETS | model with carbon tax and ETS | model with EPS | model with carbon tax, ETS and EPS |
| previous year lnCO2 per capita | 0.7205*** | 0.7233*** | 0.7191*** | 0.7471*** | 0.7320*** |
| | 0.0542 | 0.0538 | 0.0543 | 0.0598 | 0.0635 |
| carbon tax (real, lagged) | -0.0013*** | | -0.0012*** | | -0.0007*** |
| | 0.0003 | | 0.0003 | | 0.0002 |
| ETS (real, lagged) | | -0.0017*** | -0.0014*** | | -0.0003 |
| | | 0.0005 | 0.0005 | | 0.0004 |
| EPS (lagged) | | | | -0.0164** | -0.0137** |
| | | | | 0.0064 | 0.0066 |
| GDP per capita (log) | 0.1884*** | 0.1853*** | 0.1899*** | 0.1712*** | 0.1824*** |
| | 0.0438 | 0.0432 | 0.0440 | 0.0575 | 0.0622 |
| GDP growth | 0.4051*** | 0.3994*** | 0.3952*** | 0.5099*** | 0.5063*** |
| | 0.0983 | 0.0980 | 0.0977 | 0.1334 | 0.1332 |
| urbanisation rate | 0.3609** | 0.3655** | 0.3517** | 0.2500* | 0.2127 |
| | 0.1633 | 0.1623 | 0.1624 | 0.1396 | 0.1353 |
| manufacturing / GDP | 0.5452*** | 0.5447*** | 0.5440*** | 0.1661 | 0.1144 |
| | 0.1400 | 0.1405 | 0.1396 | 0.1875 | 0.2002 |
| share of electricity from oil | 0.1571*** | 0.1529*** | 0.1565*** | 0.1387** | 0.1575** |
| | 0.0397 | 0.0392 | 0.0397 | 0.0564 | 0.0587 |
| share of electricity from coal | 0.2414*** | 0.2355*** | 0.2335*** | 0.2331*** | 0.2618*** |
| | 0.0734 | 0.0721 | 0.0720 | 0.0642 | 0.0740 |
| share of electricity from renewables | -0.1872** | -0.1874** | -0.1640** | -0.2361** | -0.2020** |
| | 0.0788 | 0.0796 | 0.0758 | 0.0858 | 0.0848 |
| observations | 3881 | 3881 | 3881 | 653 | 653 |
| number of countries | 121 | 121 | 121 | 30 | 30 |
| R2 within | 0.822 | 0.822 | 0.822 | 0.879 | 0.882 |
| R2 between | 0.989 | 0.989 | 0.989 | 0.971 | 0.97 |

Note: Sample period: from 1971 to 2016, annual data. Cluster-robust standard errors reported in brackets are clustered at the country level. \*\*\*/\*\*/\* denote statistical significance at the 1%/5%/10% level. (I): real price of first carbon tax in USD/tCO2 equivalents (deflated by US CPI); (II): real price of ETS in USD/tCO2 equivalents (deflated by US CPI); (IV): EPS index.

Finally, we exclude countries whose economic development, as measured by GDP per capita, is below that of the country with the lowest GDP per capita among those that have implemented carbon taxes or ETS. We do so by restricting GDP per capita to above $1500 in constant 2010 dollars in columns I to III of Appendix Table A1. Very similar results are obtained when this restriction is imposed.

## 6 Conclusion

We use a comprehensive database of 121 countries to study how climate policies have affected carbon dioxide emissions ex-post, with data on carbon emissions and macroeconomic variables over the 1971-2016 period and data on climate policies. As climate policies we consider carbon taxes and emissions trading systems, as well as a broad index for the stringency of climate policies. Overall, we find that higher carbon taxes and prices of permits in the ETS are associated with reduced carbon emissions. Further, more stringent climate policies as measured by a broader index for OECD and major emerging economies, have also significantly reduced carbon emissions.

Overall, an increase in carbon taxes by $10/tCO_2$ reduces $CO_2$ emissions per capita by 1.3% in the short run and by 4.6% in the long run. This negative effect on emissions is statistically significant for all nine specifications that were used, with *p-values* that are always below 0.01. The same increase in the prices of ETS permits reduces $CO_2$ emissions per capita by 1.4% in the short run and 5.0% in the long run, although this effect is found to be less robust to alternative specifications. More stringent climate policies as measured by a broad index for OECD and major emerging economies also significantly reduce carbon emissions, with an increase of one standard deviation in the index reducing $CO_2$ emissions per capita by around 1.5% in the short run and 6.0% in the long run.

Our findings are relevant for the design of climate policies. The fact that higher carbon tax rates and prices of permits in the ETS rates have reduced carbon emissions suggests that further increases in these and expansion to more countries, thus covering a greater share of global carbon emissions, are promising avenues to speed up the necessary transition towards much lower carbon emissions economies. Further, the finding that more broadly, stringent climate policies have reduced carbon emissions, indicates that future enhancements in a wider range of climate policies can also be helpful for a speedy transition towards much lower carbon emissions, ideally to net zero emissions.

# Appendix

**Table A1**
**Effects of climate policies on CO2 emissions: restricted sample**

Dependent variable: lnCO2 emissions per capita (in metric tons, log)

| | model | | | | |
|---|---|---|---|---|---|
| | (I) | (II) | (III) | (IV) | (V) |
| | model with carbon tax | model with ETS | model with carbon tax and ETS | model with EPS | model with carbon tax, ETS and EPS |
| previous year lnCO2 per capita | 0.6963*** | 0.7010*** | 0.6948*** | 0.7471*** | 0.7334*** |
| | 0.0769 | 0.0762 | 0.0769 | 0.0598 | 0.0627 |
| carbon tax (lagged) | -0.0012*** | | -0.0011*** | | -0.0006*** |
| | 0.0003 | | 0.0003 | | 0.0002 |
| ETS (lagged) | | -0.0014*** | -0.0010** | | -0.0003 |
| | | 0.0005 | 0.0005 | | 0.0004 |
| EPS (lagged) | | | | -0.0164** | -0.0135** |
| | | | | 0.0064 | 0.0066 |
| GDP per capita (log) | 0.1479*** | 0.1444*** | 0.1493*** | 0.1712*** | 0.1801*** |
| | 0.0446 | 0.0438 | 0.0449 | 0.0575 | 0.0616 |
| GDP growth | 0.3347*** | 0.3285*** | 0.3253*** | 0.5099*** | 0.5012*** |
| | 0.0968 | 0.0976 | 0.0969 | 0.1334 | 0.1334 |
| urbanisation rate | 0.4650** | 0.4695** | 0.4566** | 0.2500* | 0.2086 |
| | 0.2168 | 0.2171 | 0.2158 | 0.1396 | 0.1352 |
| manufacturing / GDP | 0.3376** | 0.3320** | 0.3396** | 0.1661 | 0.1036 |
| | 0.1419 | 0.1410 | 0.1424 | 0.1875 | 0.2038 |
| share of electricity from oil | 0.1547*** | 0.1477*** | 0.1547*** | 0.1387** | 0.1614** |
| | 0.0502 | 0.0490 | 0.0502 | 0.0564 | 0.0592 |
| share of electricity from coal | 0.3375*** | 0.3276*** | 0.3300*** | 0.2331*** | 0.2640*** |
| | 0.0989 | 0.0972 | 0.0976 | 0.0642 | 0.0747 |
| share of electricity from renewables | -0.1006 | -0.1054 | -0.0816 | -0.2361** | -0.1975** |
| | 0.0861 | 0.0858 | 0.0848 | 0.0858 | 0.0852 |
| observations | 2875 | 2875 | 2875 | 653 | 653 |
| number of countries | 102 | 102 | 102 | 30 | 30 |
| R2 within | 0.842 | 0.841 | 0.842 | 0.879 | 0.881 |
| R2 between | 0.978 | 0.979 | 0.978 | 0.971 | 0.971 |

Note: Sample period: from 1971 to 2016, annual data. Cluster-robust standard errors reported in brackets are clustered at the country level. ***/**/* denote statistical significance at the 1%/5%/10% level. (I): nominal price of first carbon tax in USD/tCO2 equivalents; (II): nominal price of ETS in USD/tCO2 equivalents; (IV): EPS index; for (I) to (III): restricted to GDP per capita in constant 2010 dollars of above 1500.

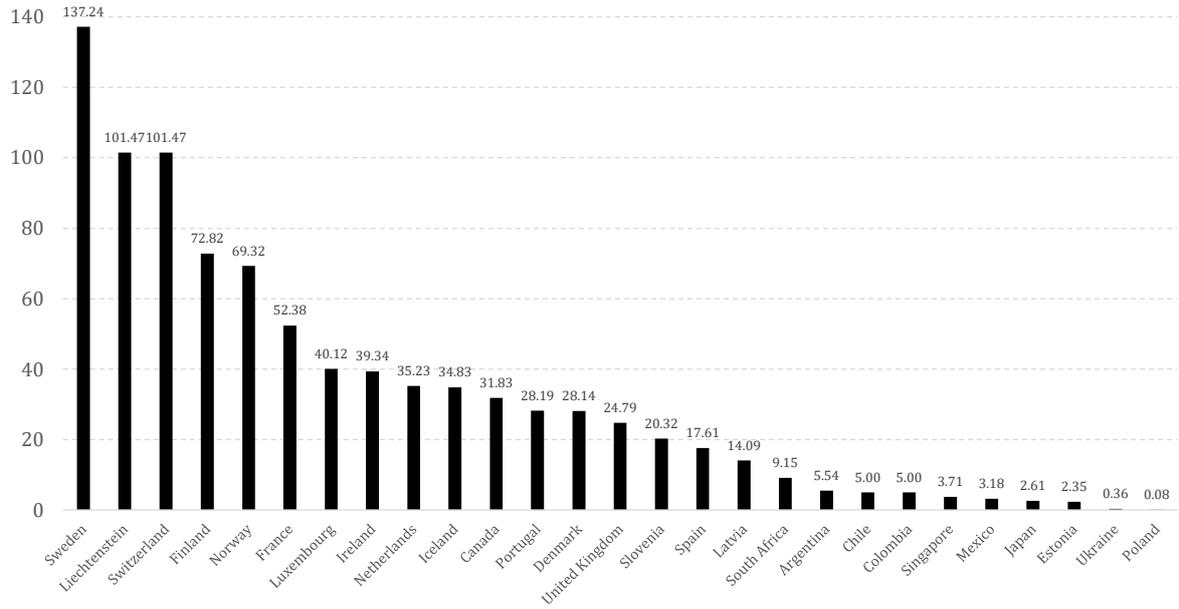

**Figure A1: Prices of carbon taxes as of 2021** ($ per tCO2e)

Notes: The figure for Canada refers to the federal fuel charge. For the United Kingdom, to the UK carbon price support.

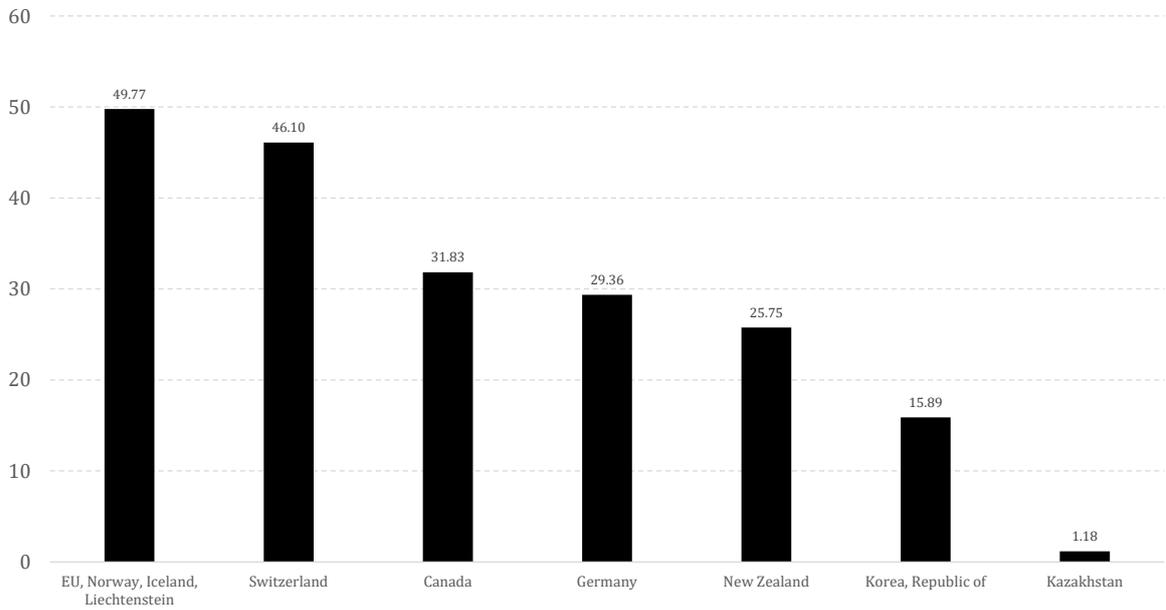

**Figure A2: Prices of carbon ETS as of 2021** ($ per tCO2e)